\documentclass[pre,showpacs,floats,twocolumn,floatfix,superscriptaddress]{revtex4-1}
\usepackage{graphics,graphicx,dcolumn,bm,fleqn,epic,eepic,float}
\usepackage{amssymb,amsmath,multirow,rotate,rotating,color}
\usepackage[utf8]{inputenc}
\usepackage{subfigure}
\newcommand{\eqnref}[1]{Eq.~(\ref{eq:#1})}
\newcommand{\RY}[1]{#1} % self-inspired changes

\definecolor{tuered}{RGB}{214,0,74}
\definecolor{tueblue}{RGB}{0,102,204}
\definecolor{grey}{RGB}{128,128,128}
\definecolor{specialgreen}{RGB}{0,214,0}

\renewcommand{\vec}[1]{\mathbf{#1}}

\usepackage{refcheck}
\norefnames
\nocitenames

\newcommand{\pca}{\theta}

\newcommand{\pr}{R}

\newcommand{\prp}{R_{\parallel}}

\newcommand{\pro}{R_{\perp}}

\newcommand{\distpc}{r}

\newcommand{\uvpc}{\mathbf{\hat{r}}_{ij}}

\newcommand{\vpc}{\mathbf{r}_{ij}}

\newcommand{\po}{\mathbf{\hat{o}}}

\newcommand{\pvf}{C}

\newcommand{\pcf}{\chi}

\newcommand{\pn}{N}

\newcommand{\pms}{m}

\newcommand{\pv}{\vec{u}_{\rm par}}

\newcommand{\pacc}{\dot{\vec{u}}_{\rm par}}

\newcommand{\Fp}{\vec{F}}

\newcommand{\factorC}{{\cal C}}

\newcommand{\mip}{J}

\newcommand{\Tp}{\vec{D}}

\newcommand{\prot}{\vec{\omega}_{\rm par}}

\newcommand{\protacc}{\dot{\vec{\omega}}_{\rm par}}

\newcommand{\opS}{S}

\newcommand{\opQ}{Q}

\newcommand{\Pcf}{g}

\newcommand{\acf}{h}

\newcommand{\pcfnorm}{\Pcf_n}

\newcommand{\ifa}{A_I}

\newcommand{\ifl}{L_I}

\newcommand{\sv}{V_S}

\newcommand{\sL}{L_S}

\newcommand{\pifa}{A_P}

\newcommand{\pa}{\vartheta}

\newcommand{\Aa}{\phi}

\newcommand{\Par}{m}

\newcommand{\st}{\sigma}

\newcommand{\dip}{\xi}

\newcommand{\dxpi}{x}

\newcommand{\dypi}{z}

\newcommand{\Doi}{y}

\newcommand{\Fsc}{\vec{F}^c}

\newcommand{\mwf}{\Psi^c}

\newcommand{\ds}{L(t)}

\newcommand{\stf}{\varsigma}

\newcommand{\stfn}{\stf_n}

\newcommand{\wsf}{k}

\newcommand{\wvsf}{\vec{k}}

\newcommand{\sfpr}{\Upsilon}

\newcommand{\spdfi}{f_i}

\newcommand{\lpx}{\vec{x}}

\newcommand{\lvci}{\vec{c}_i}

\newcommand{\dens}{\rho}

\newcommand{\densb}{\rho^b}

\newcommand{\densr}{\rho^r}

\newcommand{\fop}{\varphi}

\newcommand{\fopft}{\tilde{\fop}}

\newcommand{\fopftfl}{\fop'}

\begin{document}

\title{Timescales of emulsion formation caused by
    anisotropic particles}

\author{Florian G\"unther}
%\email{f.s.guenther@tue.nl}
\affiliation{Department of Applied Physics, Eindhoven University of Technology, Den Dolech 2, NL-5600MB Eindhoven, The Netherlands}

\author{Stefan Frijters}
%\email{s.c.j.frijters@tue.nl}
\affiliation{Department of Applied Physics, Eindhoven University of Technology, Den Dolech 2, NL-5600MB Eindhoven, The Netherlands}

\author{Jens Harting} 
\email{j.harting@tue.nl}
\affiliation{Department of Applied Physics, Eindhoven University of Technology, Den Dolech 2, NL-5600MB Eindhoven, The Netherlands}
\affiliation{Institute for Computational Physics, University of Stuttgart,
  Allmandring 3, D-70569 Stuttgart, Germany} 

%\date{\today \revision}
\date{\today}

\begin{abstract}
Particle stabilized emulsions have received an enormous interest in the
recent past, but our understanding of the dynamics of emulsion formation
is still limited. For simple spherical particles, the time dependent
growth of fluid domains is dominated by the formation of droplets,
particle adsorption and coalescence of droplets (Ostwald ripening), which
eventually can be almost fully blocked due to the presence of the particles.
Ellipsoidal particles are known to be more efficient stabilizers of fluid
interfaces than spherical particles and their anisotropic shape and the
related additional rotational degrees of freedom have an impact on the
dynamics of emulsion formation. In this paper, we investigate this point
by means of simple model systems consisting of a single ellipsoidal
particle or a particle ensemble at a flat interface as well as a particle
ensemble at a spherical interface. By applying combined multicomponent
lattice Boltzmann and molecular dynamics simulations we demonstrate that
the anisotropic shape of ellipsoidal particles causes two additional
timescales to be of relevance in the dynamics of emulsion formation: a
relatively short timescale can be attributed to the adsorption of single
particles and the involved rotation of particles towards the interface. As soon
as the interface is jammed, however, capillary interactions between the
particles cause a local reordering on very long timescales leading to a
continuous change in the interface configuration and increase of interfacial
area. This effect can be utilized to counteract the thermodynamic instability
of particle stabilized emulsions and thus offers the possibility to produce
emulsions with exceptional stability. 
\end{abstract}


\maketitle

\section{Introduction}
\label{sec:introduction}
Particle stabilized emulsions play an important role in pharmaceutical, food,
oil and cosmetic industries~\cite{Dickinson2010a}. The particles are adsorbed
at the interface between two immiscible fluids and as such stabilize the
emulsion. The stability of the emulsions depends on several parameters like
particle coverage at the interfaces and the wettability of the particles. It
was found that the particle coverage at the interface is the most important
parameter for stabilizing emulsions~\cite{Fan2012a}. The colloidal
particles act in a similar way as surfactants. In both cases the free energy of
the interface is reduced. However, the fluid-fluid interfacial tension is not
being modified by particles~\cite{Frijters2012a}.

Several types of particle stabilized emulsions are known including the
bicontinuous interfacially jammed emulsion gel (bijel) and the more widely
known Pickering emulsion. The Pickering emulsion was discovered in the
beginning of the 20\textsuperscript{th} century independently by Pickering and
Ramsden~\cite{Pickering1907a,Ramsden1903a}. It consists of discrete particle
covered droplets of a fluid immersed in a second fluid.  The bijel was
predicted in 2005 by simulations and experimentally realized for the first time
in 2007~\cite{Stratford2005a,Herzig2007a}. It consists of two continuous
phases. The choice of control parameters such as particle concentration,
particle wettability and ratio between the two fluids determines if a bijel or
a Pickering emulsion is obtained~\cite{Guenther2012a,Jansen2011a}. There are
many kinds of particles/colloid types which can stabilize an emulsion. I.e.,
next to spheres~\cite{He2007a,Aveyard2003a}, the colloidal particles can also
be of more complex nature and include anisotropic
shapes~\cite{Kalashnikova2013a}, magnetic
interactions~\cite{Kim2010a,Melle2005a}, or anisotropic Janus style
properties~\cite{Binks2001a}.

The influence of the particle shape on the stabilization of Pickering emulsions
was studied experimentally with prolate and oblate ellipsoids, e.g. in
Ref.~\cite{Madivala2009b}. As the degree of the particle anisotropy increases,
the effective coverage area increases. In this way they are more efficient
stabilizers for emulsions than spherical particles. Furthermore, the
rheological properties of the emulsion vary with changing aspect ratio because
the coverage of the fluid interfaces and the capillary interactions differ.

In Refs.~\cite{Guenther2012a,deGraaf2010a,Dong2005a,Bresme2007b,Faraudo2003a} the
adsorption of a single particle at a flat interface is studied in absence of
external fields. The stable configuration for elongated ellipsoids is the
orientation parallel to the interface~\cite{Guenther2012a}. This state
minimizes the free energy of the particle at the interface by reducing the interfacial
area~\cite{deGraaf2010a,Bresme2007b,Faraudo2003a}. If the particle shape is more complex like e.g. the
super-ellipsoidal hematite particle~\cite{Morgan2013a}, several equilibrium
orientations are possible. 

Furthermore, if particles are adsorbed at an interface they generally deform
the interface. This deformation can be caused for example by particle
anisotropy~\cite{Lehle2008a}, external forces such as gravity or
electromagnetic forces acting on the particles~\cite{Bleibel2011b,Bleibel2013},
or non-constant interface curvature~\cite{Zeng2012a}. This deformation leads to
capillary interactions between the particles. In case of ellipsoids at a flat
interface it is a quadrupolar potential~\cite{Botto2012b}, which leads to
spatial ordering~\cite{Madivala2009a}.

In general, particle stabilized emulsions are thermodynamically unstable and
just kinetically stable. The energetic penalty for creating the interface is
much higher than the entropic increase. While thermodynamic stability for
emulsions has been reported in some special cases, one can generally assume
that this requires the interplay of several effects such as particle
interactions due to charges, amphiphilic interactions (Janus particles) or
additional degrees of freedom~\cite{Sacanna2007b,Kegel2009a,Aveyard2012}.

Due to the short timescales and limited optical accessibility, the dynamics of
the formation of emulsions has only found limited attendance so
far~\cite{Dai2008}. The focus of the current article is to study the influence
of the geometrical anisotropy and rotational degrees of freedom of ellipsoidal
particles on the time development of fluid domain sizes in particle-stabilized
emulsions. To obtain a deeper understanding of the individual contributions to
the stabilization and formation process due to the particles we investigate
model systems involving either a single particle or particle ensembles at a
simple interface. We will demonstrate that the rotational degrees of freedom of
ellipsoids can have an impact on the domain growth and might be a suitable way
to generate particle stabilized emulsions with exceptional long-term stability.

This article is organized as follows: the simulation method is
introduced in section II. Dynamic emulsion properties are studied in section
III.  Sections IV and V discuss a single particle and a particle ensemble at a
flat interface, respectively. Section VI describes the behavior of a particle
ensemble at a spherical interface. We finalize the paper with a conclusion.

\section{Simulation method}
\label{sec:sim-method}
\subsection{The lattice Boltzmann method}
\label{ssec:lb}
For the simulation of the fluids the lattice Boltzmann method is
used~\cite{succi2001a}.
The discrete form of the Boltzmann equation can be written
as~\cite{Frijters2012a}
\begin{equation}
  \label{eq:LBG}
  \spdfi^c(\lpx + \lvci \Delta t , t + \Delta t)=\spdfi^c(\lpx,t)+\Omega_i^c(\lpx,t)
  \mbox{,}
\end{equation}
where $\spdfi^c(\lpx,t)$ is the single-particle distribution function for fluid
component $c$ with discrete lattice velocity $\lvci$ at time $t$ located
at lattice position $\lpx$. The D3Q19 lattice with the lattice constant
$\Delta x$ for three dimensions and with nineteen velocity
directions is used. $\Delta t$ is the
timestep and 
\begin{equation}
  \label{eq:BGK_collision_operator}
  \Omega_i^c(\lpx,t) = -\frac{\spdfi^c(\lpx,t)- \spdfi^\mathrm{eq}(\rho^c(\lpx,t), \vec{u}^c(\lpx,t))}{\left( \tau^c / \Delta t \right)}
\end{equation}
is the Bhatnagar-Gross-Krook (BGK) collision
operator~\cite{bhatnagar1954a}. The density is defined as
%$\rho^c(\lpx,t)=\sum_i\spdfi^c(\lpx,t)=\Delta m/(\Delta x)^3$. $\Delta m$ is a
%unit mass.
$\rho^c(\lpx,t)=\rho_0\sum_i\spdfi^c(\lpx,t)$ where $\rho_0$ is the
proportionality factor of the density.
$\tau^c$ is the relaxation time for the component $c$ and
\begin{multline}
  \label{eq:equilibrium-distribution}
  \spdfi^{\mathrm{eq}}(\rho^c,\vec{u}^c) = \zeta_i \rho^c \bigg[ 1 + \frac{\lvci \cdot \vec{u}^c}{c_s^2} + \frac{ \left( \lvci \cdot \vec{u}^c \right)^2}{2 c_s^4} \\ - \frac{ \left( \vec{u}^c \cdot \vec{u}^c \right) }{2 c_s^2} + \frac{ \left( \lvci \cdot \vec{u}^c \right)^3}{6 c_s^6} - \frac{ \left( \vec{u}^c \cdot \vec{u}^c \right) \left( \lvci \cdot \vec{u}^c \right )}{2 c_s^4} \bigg]
  \mbox{}
\end{multline}
is the third order equilibrium distribution function. 
\begin{equation}
  \label{eq:sos}
  c_s = \frac{1}{\sqrt{3}} \frac{\Delta x}{\Delta t}
  \mbox{}
\end{equation}
is the speed of sound, $\vec{u}^c=\sum_i\spdfi^c(\lpx,t)\lvci/\rho^c(\lpx,t)$ is the velocity and $\zeta_i$ is a coefficient depending on the
direction: $\zeta_0=1/3$ for the zero velocity, $\zeta_{1,\dots,6}=1/18$ for
the six nearest neighbors and $\zeta_{7,\dots,18}=1/36$ for the next nearest
neighbors in diagonal direction.
The kinematic viscosity can be calculated as
\begin{equation}
  \label{eq:kinvis}
  \nu^c = c_s^2 \Delta t \left( \frac{\tau^c}{\Delta t} - \frac{1}{2} \right)
  \mbox{.}
\end{equation}
In the following we choose $\Delta x = \Delta t = \rho_0 = 1$ for
simplicity.  In all simulations the
relaxation time is set to $\tau^c \equiv 1$.

\subsection{Multicomponent lattice Boltzmann}
\label{ssec:multicomponent-lb}
There are different extensions for the lattice Boltzmann method to simulate
multi-component and multiphase
systems~\cite{shan1993a,Orlandini1995a,Swift1996a,Lishchuk2003a,LeeFischer06}. An overview
on different methods for multi-component fluid systems and the 
treatment of fluid-fluid interfaces is given in Ref.~\cite{Krueger2012c}. In
this paper, the method introduced by Shan and Chen is used~\cite{shan1993a}. Every species
has its own distribution function following \eqnref{LBG}. To obtain an interaction
between the different components a force
\begin{equation}
  \label{eq:sc}
  \Fsc(\lpx,t) = -\mwf(\lpx,t) \sum_{c'}g_{cc'} \sum_{\lpx'} \Psi^{c'}(\lpx',t) (\lpx'-\lpx)
  \mbox{}
\end{equation} 
is calculated locally and is included
in the equilibrium distribution function.
it is summed up over the different fluid species $c'$ and $\lpx'$, the
nearest neighbors of lattice positions $\lpx$.
$g_{cc'}$ is the coupling constant between the species and
$\mwf(\mathbf{x},t)$ is a monotonous weight function representing an
effective mass. For the results presented here, the form
\begin{equation}
  \label{eq:psifunc}
  \mwf(\lpx,t) \equiv \Psi(\rho^c(\lpx,t) ) = 1 - e^{-\rho^c(\lpx,t)}
\end{equation}
is used. To incorporate $\Fsc(\lpx,t)$ in $\spdfi^\mathrm{eq}$ we define
\begin{equation}
  \label{eq:delta-u}
  \Delta \vec{u}^c(\lpx,t) = \frac{\tau^c \Fsc(\lpx,t)}{\rho^c(\lpx,t)}
  \mbox{.}
\end{equation}
The macroscopic velocity included in $\spdfi^\mathrm{eq}$ is shifted by $\Delta \vec{u}^c$ as
\begin{equation}
  \label{eq:delta-u-bulk}
  \vec{u}^c(\lpx,t) = \frac{\sum_i f^c_i(\lpx,t) \lvci}{\rho^c(\lpx,t)} - \Delta \vec{u}^c(\lpx,t)
  \mbox{.}
\end{equation}
As we are interested in immiscible fluids we choose a positive value for
$g_{cc'}$ which leads to a repulsive interaction. This
interaction has to be strong enough to obtain two separate phases but it should not be too
high in order to keep the simulation stable. Here, we use the range of
$0.08\le g_{cc'}\le 0.14$.

\subsection{Nanoparticles}
\label{ssec:nanoparticles}
Particles are simulated with molecular dynamics where Newton's equations of
motion
\begin{equation}
  \label{eq:MD}
  \Fp=\pms\pacc
  \mbox{ and }
  \Tp=\mip\protacc
\end{equation}
are solved by a leap frog integrator. $\Fp$ and $\Tp$ are the force and torque
acting on the particle with mass $\pms$ and moment of inertia $\mip$. $\pv$ and
$\prot$ are the velocity and the rotation vector of the particle.\\ The particles are
also discretized on the lattice. They are coupled to both fluid species by a
modified bounce-back boundary condition which was originally introduced by
Ladd~\cite{Jansen2011a,aidun1998a,ladd1994a,ladd1994b,ladd2001a}. This changes the lattice
Boltzmann equation as follows:
\begin{equation}
  \label{eq:mbb}
  \spdfi^c(\lpx+\lvci,t+1) = f^c_{\bar{i}}(\lpx+\lvci,t) + \Omega_{\bar{i}}^c(\lpx+\lvci,t) + \factorC
  \mbox{,}
\end{equation}
where $\vec{c}_i$ is the velocity vector pointing to the next neighbor.
$\factorC$ depends linearly on the local particle velocity, $\bar{i}$ is defined in a
way that $\lvci = -\vec{c}_{\bar{i}}$ is fulfilled.
A change of the fluid momentum due to a particle leads to a change of the
particle momentum in order to keep the total momentum conserved:
\begin{equation}
  \label{eq:mbb2}
  \Fp(t) = \big( 2f_{\bar{i}}^c(\lpx+\lvci,t) + \factorC \big) \vec{c}_{\bar{i}}
  \mbox{.}
\end{equation}
If the particle moves, some lattice nodes become free and others become
occupied. The fluid on the newly occupied nodes is deleted and its momentum is
transferred to the particle as
\begin{equation}
  \label{eq:mom-transfer-to-ptcl}
  \vec{F}(t) = - \sum_c \rho^c(\lpx,t) \vec{u}^c(\lpx,t)
  \mbox{.}
\end{equation}
A newly freed node (located at $\lpx$) is filled with the average density of the $N_{\mathrm{FN}}$ neighboring fluid
lattice nodes $\lpx_{i_\mathrm{FN}}$ for each component $c$,
\begin{equation}
  \label{eq:rho-surr}
  \overline{\rho}^c(\lpx,t) \equiv \frac{1}{N_{\mathrm{FN}}} \sum_{i_\mathrm{FN}}  \rho^c(\lpx+\vec{c}_{i_{\mathrm{FN}}},t)
  \mbox{.}
\end{equation}
Hydrodynamics leads to a lubrication force between the particles. This force is reproduced automatically by the simulation for sufficiently
large particle separations. If the
distance between the particles is so small that no free lattice point exists between
them this reproduction fails.
If the smallest distance between two identical spheres with radius $R$ is
smaller than a critical value $\Delta_c= \frac{2}{3}$ the correction term is given as~\cite{ladd2001a}:
\begin{equation}
  \label{eq:lubrication}
\mathbf{F}_{ij}=\frac{3\pi\mu\RY{R^2}}{2}\uvpc(\uvpc(\mathbf{u}_i-\mathbf{u}_j))\left(\frac{1}{r_{ij}-2R}-\frac{1}{\Delta_c}\right)
\mbox{.}
\end{equation}
$\mu$ is the dynamic viscosity,
%$\Delta_c$ the cutoff length which is chosen to be $\Delta_c=2/3$,
$\uvpc$ a unit vector pointing from one particle center to the other one and
$\mathbf{u}_i$ is the velocity of particle $i$.
To use this potential for
ellipsoidal particles \eqnref{lubrication} is generalized in a way proposed by Berne
and Pechukas~\cite{berne1972a,Janoschek2010b,Guenther2012a}. We define $\sigma=2\pr$ and
$\epsilon=\frac{3\pi\mu}{8}\sigma$. Both are extended to the anisotropic case
as
\begin{equation}
\begin{split}
\epsilon(\po_i,\po_j)=\frac{\overline{\epsilon}}{\sqrt{1-\sfpr^2(\po_i\po_j)^2}}
\quad\mbox{and}\quad\quad\quad\quad\quad\quad\quad\quad\quad
\\
\label{eq:sigmaaniso}
\sigma(\po_i,\po_j,\uvpc)
=\frac{\overline{\sigma}}{\sqrt{1-\frac{\sfpr}{2}(\frac{(\uvpc\po_i+\uvpc\po_j)^2}{1+\sfpr\po_i\po_j}+\frac{(\uvpc\po_i-\uvpc\po_j)^2}{1-\sfpr\po_i\po_j})}}
\mbox{,}
 \end{split}
\end{equation}
with $\overline{\sigma}=2\pro$,
$\overline{\epsilon}=\frac{3\pi\mu}{8}\overline{\sigma}$,
$\sfpr=\frac{\prp^2-\pro^2}{\prp^2+\pro^2}$ and $\po_i$ the
orientation unit vector of particle $i$. $\prp$ and $\pro$ are the parallel and the
orthogonal radius of the ellipsoid. Using \eqnref{sigmaaniso} we can rewrite
\eqnref{lubrication} and obtain
\begin{equation}
  \label{eq:lubricationres}
\mathbf{F}_{ij}(\po_i,\po_j,\vpc)=\epsilon(\po_i,\po_j)\mathbf{\tilde{F}}_{ij}\left(\frac{r_{ij}}{\sigma(\po_i,\po_j,\uvpc)}\right)
\mbox{.}
\end{equation}
$\mathbf{\tilde{F}}$ is a dimensionless function taking the specific form of
the force into account and in this example it is $\mathbf{\tilde{F}}(r)=\uvpc(\uvpc(\mathbf{u}_i-\mathbf{u}_j))(\frac{1}{r-1}-\frac{\sigma}{\Delta_c})$.\\
The lubrication force (including the correction) already reduces the
probability that the particles come closely together and overlap. For the few
cases where the particles still would overlap we introduce
the direct potential between the particles which is assumed to be a hard core
potential. To approximate the hard core potential we use the Hertz
potential~\cite{Hertz1881a} which has the following shape for two identical
spheres with radius $R$:
\begin{equation}
  \label{eq:Hpot}
\phi_H=K_H(2\pr-\distpc)^{5/2}\mbox{ for }\distpc<2\pr
  \mbox{.}
\end{equation}
$\distpc$ is the distance between particle centers.
For larger distances $\phi_H$ vanishes. $K_H$ is a force constant and is
chosen to be $K_H=100$ for all simulations. To use this potential for
ellipsoidal particles \eqnref{Hpot} is generalized in a similar way as the
lubrication force. Using \eqnref{sigmaaniso}, $\sigma=2\pr$ and
$\epsilon=K_H\sigma^{\frac{5}{2}}$ we can rewrite
\eqnref{Hpot} and obtain
\begin{equation}
\phi_H(\po_i,\po_j,\mathbf{r}_{ij})
=\epsilon(\po_i,\po_j)\tilde{\phi}_H \left(\frac{r_{ij}}{\sigma(\po_i,\po_j,\uvpc)}\right)
\mbox{.}
\end{equation}
$\tilde{\phi}_H$ is a dimensionless function taking the specific form of
the potential into account and in this example it is $\tilde{\phi}_H(x)=(1-x)^{5/2}$.\\
%\begin{equation}
%\mathbf{F}_{ij}=\frac{3\pi\mu\RY{R^2}}{2}\uvpc(\uvpc(\mathbf{u}_i-\mathbf{u}_j))\left(\frac{1}{r_{ij}-2R}-\frac{1}{\Delta_c}\right)
%\end{equation}
The Shan-Chen forces also act between a node in the outer shell of a particle
and its neighboring node outside of the particle. This would lead to an increase
of the fluid density around the particle.
Therefore, the nodes in the outer shell of the particle are filled with a virtual fluid
corresponding to the average of the value in the neighboring free nodes for
each fluid component:
$\rho_{\mathrm{virt}}^c(\lpx,t) = \overline{\rho}^c(\lpx,t)$.
This can be used to control the wettability properties of the particle
surface for the special case of two fluid species which will be named red and blue. We define the parameter $\Delta\rho$ and call it particle color. For
positive values of $\Delta\rho$ we add it to the red fluid component: 
\begin{equation}
  \label{eq:red-colour}
  \rho_{\mathrm{virt}}^r=\overline{\rho}^r+\Delta\rho
  \mbox{.}
\end{equation}
For negative values we add its absolute value to the blue component:
\begin{equation}
  \label{eq:blue-colour}
  \rho_{\mathrm{virt}}^b=\overline{\rho}^b+|\Delta\rho|
  \mbox{.}
\end{equation}
In Ref.~\cite{Guenther2012a} it is shown that there is a linear relation between
$\Delta\rho$ and the three-phase contact angle $\pca$.

\section{Emulsions}
\begin{figure}
\centering
\includegraphics[width=0.52\textwidth]{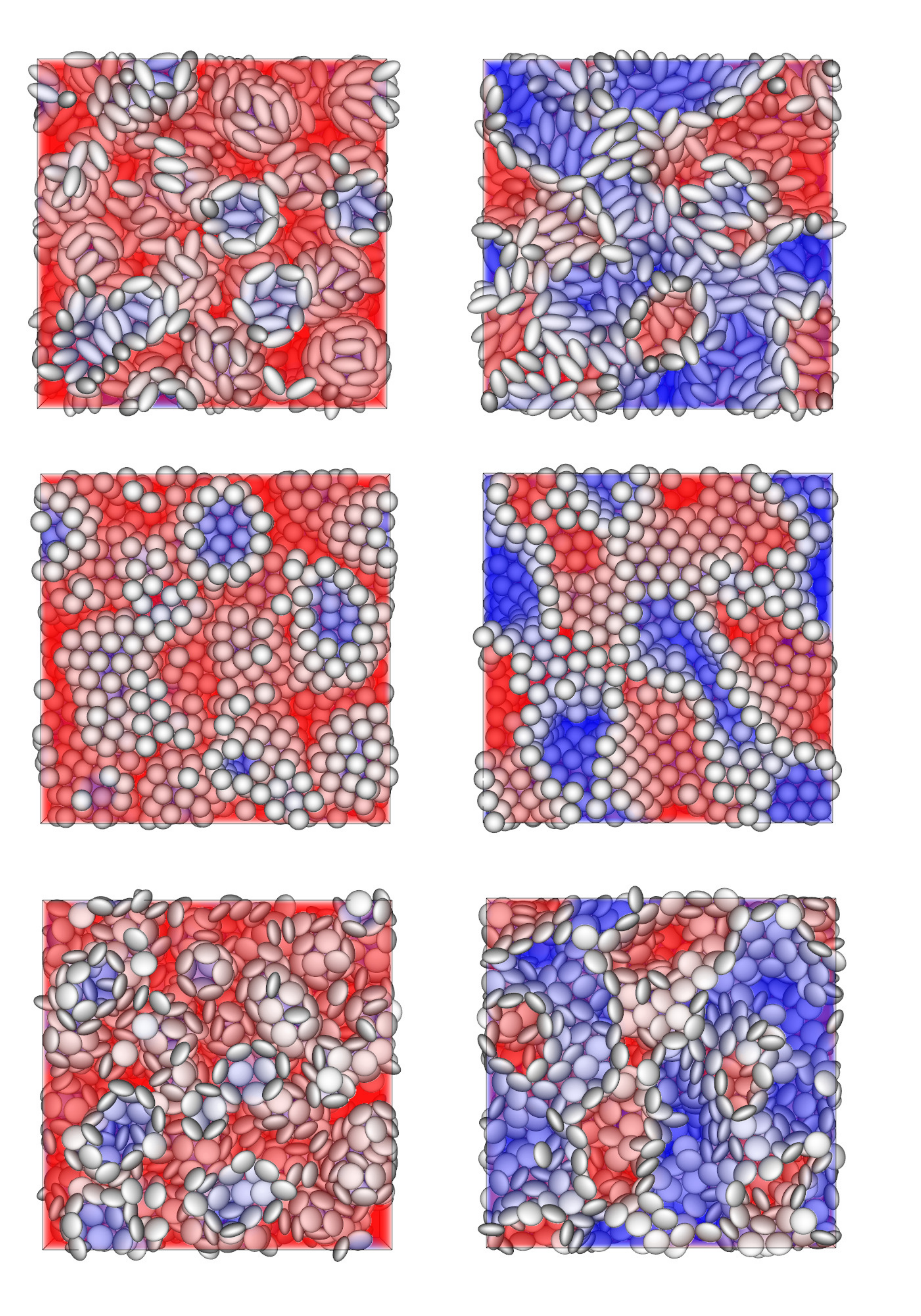}
\caption{Snapshots of typical simulated Pickering emulsions (left) and bijels (right) after 10$^5$ timesteps. The emulsions are stabilized by prolate ellipsoids ($m=2$, top), spheres ($m=1$, center) and
oblate ellipsoids ($m=1/2$, bottom). The parameter determining if one obtains a bijel or a Pickering emulsion is the fluid ratio which is chosen as 1:1 for the bijels and 5:2 for Pickering emulsions.
}
\label{PEBijelSnapshot}
\end{figure}
In this section the different types of particle stabilized emulsions and the
effect of the particle shape on some of their properties are discussed. We find
two different types of emulsions in our simulations, namely the Pickering
emulsion (Fig.~\ref{PEBijelSnapshot}, left) and the bijel
(Fig.~\ref{PEBijelSnapshot}, right).  The choice of parameters (such as
particle contact angle, particle concentration, fluid-fluid ratio, particle
aspect ratio) determines the type of emulsions. Parameter studies for emulsions
have been discussed in Refs.~\cite{Jansen2011a} and~\cite{Guenther2012a} for
spherical and ellipsoidal particles, respectively. In the current publication
we limit ourselves to anisotropy effects on the time dependence of the emulsion
formation. We use the following particle shapes ($m=\prp/\pro$ is the particle
aspect ratio. $\prp$ and $\pro$ are the parallel and orthogonal radius of the
particles, respectively): prolate ellipsoids ($m=2$;
Fig.~\ref{PEBijelSnapshot}, top), spheres ($m=1$; Fig.~\ref{PEBijelSnapshot},
center) and oblate ellipsoids ($m=1/2$; Fig.~\ref{PEBijelSnapshot}, bottom).
For $m=1/2$ we choose $\prp=5\Delta x$ and $\pro=10\Delta x$.  For the other
values of $m$ the radii $\prp$ and $\pro$ are chosen as such that the particle
volume is kept constant, resulting in $\prp\approx12.6\Delta x$ and
$\pro\approx6.3\Delta x$ for $m=2$ as well as $\prp=\pro\approx7.9\Delta x$ for
spheres. The interaction parameter between the fluids (see \eqnref{sc}) is
chosen as $g_{br}=0.08$ which corresponds to a fluid-fluid interfacial tension
of $\st=0.0138$.  The particles are neutrally wetting (contact angle
$\pca=90^\circ$) and the particle volume concentration is chosen as
$\pvf=0.24$.  The simulated systems of volume $\sv=\sL^3$ have periodic
boundary conditions in all three directions and a side length of $\sL=256\Delta
x=32\prp$.  Initially, the particles are distributed randomly.  At each lattice
node a random value for each fluid component is chosen so that the designed
fluid-fluid ratio is kept (1:1 for the bijels and 5:2 for the Pickering
emulsions).  When the simulation evolves in time, the fluids separate and
droplets/domains with a majority of red or blue fluid form.

The average size of droplets/domains $\ds$ can be determined by measuring
\begin{equation}
  \label{eq:L}
\ds=\frac{1}{3}\sum_{i=x,y,z}\ds_i
  \mbox{.}
\end{equation}
Here,
\begin{equation}
  \label{eq:Li}
\ds_i=\frac{2\pi}{\sqrt{\langle\wsf_i^2(t)\rangle}}
\end{equation}
is the average domain size in direction $i$.
$\langle\wsf_i^2(t)\rangle=\sum_{\wvsf}\wsf_i^2(t)\stf(\wvsf,t)/\sum_{\wvsf}\wsf_i^2(t)$
is the second-order moment of the three-dimensional structure function
$\stf(\wvsf,t) = (1/\stfn)
|\fopftfl_\wvsf(t)|$. $\fopftfl=\fopft-\langle\fopft\rangle$ is the
fluctuation of $\fopft$ which is the Fourier transform of the order parameter
field $\fop=\densr-\densb$. In this publication, the time is given in
  simulation timesteps, which can be converted to physical units.
We use \eqnref{sos} and \eqnref{kinvis} to relate the
kinematic viscosity to $\Delta x$ and $\Delta t$. By assuming
$\nu=10^{-6}m^2/s$, the kinematic viscosity for water, $R=125nm$ and
$R=7.9\Delta x$ (this is the value used for the spherical particle, see above) we fix the chosen resolution of the simulation. Thus, we obtain
$\Delta x=15.8nm$ and $\Delta t=4.2\times10^{-11}s$ and a total system
size of $L_S\approx4\mu$. The interfacial tension is then $\sigma=3.14\times10^{-08}N/m$. Larger system sizes can be reached with the same computational effort by compromising on the resolution.
\begin{figure}
\centering
\includegraphics[width=0.4\textwidth,angle=0]{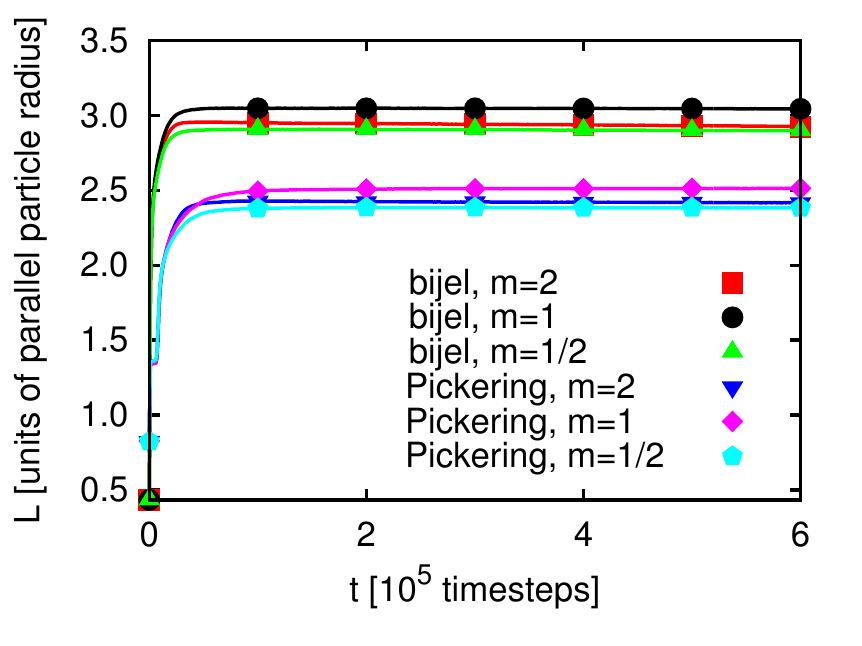}
\caption{Pickering emulsion and bijel: Time development of the average domain
size $\ds$ (see \eqnref{L}) for $m=1$ and $m=2$. At first view, a steady state
is reached after about $10^5$ timesteps. $\ds$ is larger for bijels than for
Pickering emulsions, which is due to the measurement being based on the Fourier
transform of the order parameter. Ellipsoids are able to stabilize larger
interface areas than spheres leading to smaller $\ds$.}
\label{LtvglKugelPEuBijel}
\end{figure}

The time development of $\ds$ for the three different particle types (prolate,
spherical and oblate ($m=2$, $1$ and $1/2$) and for Pickering emulsions and
bijels is shown in Fig.~\ref{LtvglKugelPEuBijel}. We can identify three
regimes: in the first few hundred timesteps the initial formation of the
droplets/domains starts. Then, the growth of droplets/domains is being driven
by Ostwald ripening. At even later times, droplets/domains grow due to
coalescence. When two droplets unify, the area coverage fraction of the
particles at the interface is increased because the surface area of the new
droplet is smaller than that of the two smaller droplets before. At some point
the area coverage fraction of the particles is sufficiently high to prevent
further coalescence. The state which is reached at that time is (at least
kinetically) stabilized and one obtains a stable emulsion. The values for
$\ds$ are larger for bijels than for Pickering emulsions. This can be
explained by the way we calculate $\ds$ (see \eqnref{L} and related text) using
a Fourier transformation of the order parameter field. 

It can clearly be seen that anisotropic particles are more efficient in
interface stabilization than spheres since they can cover larger interfacial
areas leading to smaller fluid domains (note that the simulation volume is kept
constant). However, the difference in $\ds$ for $\Par=2$ and $\Par=1/2$ is
small. This can be understood as follows: if a neutrally wetting prolate
ellipsoid is adsorbed at a flat interface, it occupies an area
$A_{P,F}(\Par>1)=\Par^{1/3}A_{p,s}$, where $A_{p,s}$ is the occupied interface
area for a sphere with the same volume.  This corresponds in the case of
$\Par=2$ to the occupied interface being larger by a factor of $1.26$ as
compared to spheres. For an oblate ellipsoid the occupied interface area is
$A_{P,F}(\Par<1)=\Par^{-2/3}A_{p,s}$ which for $\Par=1/2$ is by a factor of
$1.59$ larger than the area occupied by spheres.  Since in emulsions the
interfaces are generally not flat, these formulae can only provide a
qualitative explanation of the behavior of $\ds$: If the
interface curvature is not neglectable anymore, we loose some of the efficiency
of interface stabilization, which is more pronounced for $\Par<1$. This
explains why the value of $\ds$ for $\Par=1/2$ is only slightly smaller than
for $\Par=2$.  

It seems that $\ds$ reaches a steady state after some $10^5$ timesteps for both
types of emulsions and for all three values of $\Par$. However, if one zooms in
one can observe that $\ds$ develops for a longer time period if the particles
have a non-spherical shape. As will be demonstrated below, the reason for this phenomenon is the additional
rotational degrees of freedom due to the particle anisotropy. Furthermore, the
time development of $\ds$ for emulsions stabilized by prolate particles requires
more time than that for the oblate ones. If a particle changes its orientation
as compared to the interface or a neighboring particle this generally changes
the interface shape. In this way the domain sizes are influenced, leading to
changes of $\ds$ -- an effect which is not observed for $m=1$.
\begin{figure}
\centering
\includegraphics[width=0.4\textwidth,angle=0]{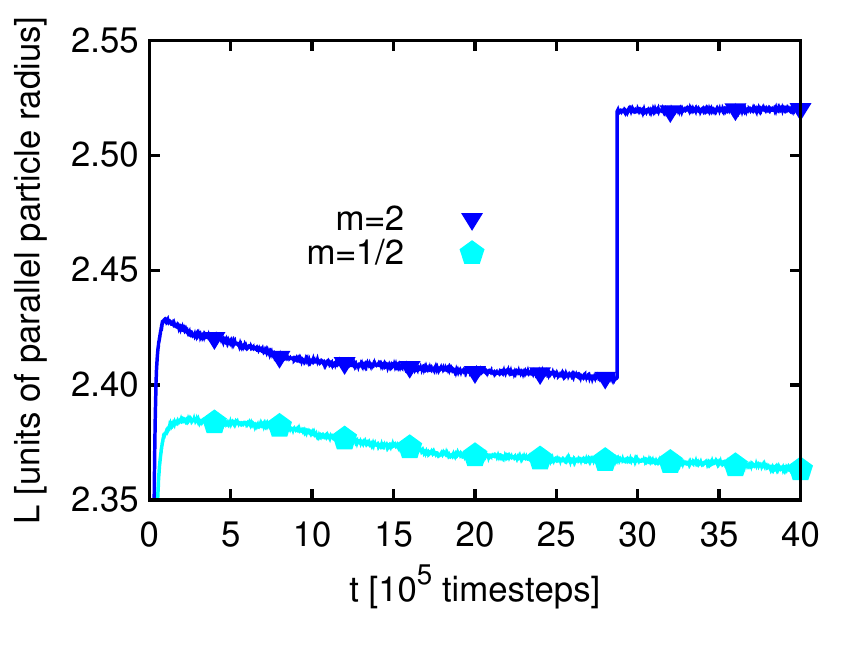}
\caption{Pickering emulsion: zoom of the time dependent average domain size $\ds$
for $m=2$, $m=1$ and $m=1/2$. The slow but continuous decrease of $\ds$ clearly
shows the occurrence of additional timescales in the domain growth. The kink in
the measurement for $m=2$ can be adhered to the coalescence of
two droplets.
% Inset: Schematic representation of the ``small'' interfaces of
%the Pickering emulsion. 
%Particle anisotropy leads to a complex time
%development of $\ds$ but the range of the variation is much smaller
%as compared to the bijel.
}
\label{dsft_zoom_PE}
\end{figure}
\begin{figure}
\centering
\includegraphics[width=0.4\textwidth,angle=0]{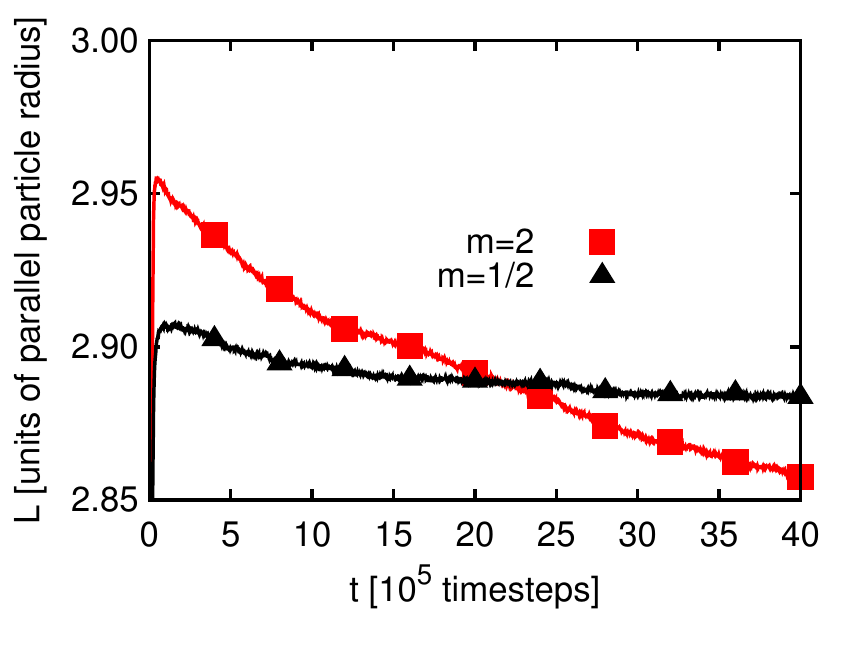}
\caption{Bijel: Zoom for $m=2$ and $m=1/2$: Time development of the average
domain sizes depicting the impact of the additional timescales. The range of
the variation of $\ds$ is larger as compared to the Pickering emulsion due to
the impact of a small deformation on the larger effective interface of the
bijel.}
\label{dsft_zoom_bijel}
\end{figure}
Fig.~\ref{dsft_zoom_PE} and Fig.~\ref{dsft_zoom_bijel} depict a zoom-in of the
time development of $\ds$ for Pickering emulsions and bijels with $m=2$ and
$m=1/2$, respectively. One observes that $\ds$ decays in all four cases. The
kink in Fig.~\ref{dsft_zoom_PE} after about 2.8 million timesteps is due to the
coalescence of two droplets of the Pickering emulsion. A substantial difference
is the range of the decay. It is larger for the bijel since it 
%In order to understand
%this phenomenon we depict in the insets of Figs.~\ref{dsft_zoom_PE} and
%\ref{dsft_zoom_bijel} a cross section of the Pickering emulsion and the bijel.
%The bijel 
consists of a single large interface whereas the Pickering emulsion
consists of many small interfaces. The large interface in the bijel is much
more deformable. This explains the larger range of the decay of $\ds$ for the
bijel. The fluctuations are of the same order for Pickering emulsions and
bijels. Furthermore, the range of the decay is larger for $m=2$ than for
$m=1/2$. The time of reordering is much shorter for $m=1/2$ as compared to
$m=2$.
These effects can be explained by the presence of additional rotational degrees
of freedom for the anisotropic particles.
While oblate particles have only a single additional rotational
degree of freedom as compared to spheres, prolate particles show an even
more complex behavior due to their second additional rotational degree of
freedom. 

In this section we demonstrated that particle anisotropy causes additional
timescales to influence the growth of domains in particle-stabilized emulsions.
In the following sections we discuss model systems in order to obtain a deeper
understanding of this effect.  We will restrict ourselves to prolate particles
with $m=2$.  Furthermore, the high resolution of the particles in the current
section was only chosen to be able to sufficiently resolve the oblate objects.
In order to reduce the required computational resources, we use smaller
particles in the model systems studied below ($\prp=8\Delta x$ and
$\pro=4\Delta x$). It has been checked carefully that the reduced particle size
does not have a qualitative impact on the results. 

%To make sure that the study is still valid, we compare the development of
%$L(t)$ for $m=2$ for the both particle sizes and for both emulsion types in
%Fig.~\ref{CompareParticleSizes}. The value of $L(t)$ is shifted by 0.5 for the
%smaller particles in order to keep them in the same range as the larger
%particles. We can see quantitative differences. System size and particle
%concentration are the same which means that the system with the smaller
%particles includes a higher number of particles. Thus a larger interface area
%can be stabilized in both emulsion types. Also the change of $L(t)$ over the
%time is larger for smaller particles. Considering this facts we can see that
%qualitative change is similar. This legitimates the use of smaller particles in
%the following.
%\begin{figure}
%\centering
%\includegraphics[width=0.4\textwidth,angle=0]{dsft_mr3_3u4zoom1vglps3.pdf}
%\caption{Bijel and Pickering emulsion (Zoom for $m=2$): Time development of the average domain sizes
%depicting the impact of the additional timescales. The development for both
%particle sizes is compared. $L(t)$ are shifted by 0.5 for the smaller
%particles. Quantitative differences but qualitative similar.}
%\label{CompareParticleSizes}
%\end{figure}

%The data for these particles are compared to $\prp=8\Delta x$ and $\pro=4\Delta
%x$ ($m=2$) corresponding to the particle in the model systems (see below) in
%order to provide quantitative comparisons. The particle size in the model
%systems discussed in the following sections is chosen to be smaller in order to
%save computing time.

%\input{1padsorption}
\section{Single particle adsorption}
\begin{figure}
\centering
\includegraphics[width=8cm]{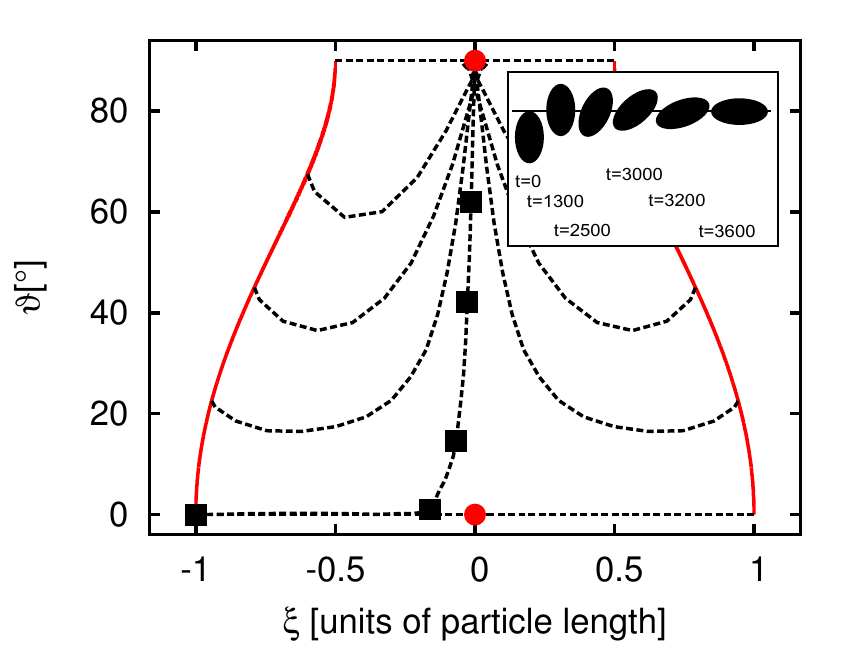}
\caption{Outer plot: $\pa$-$\xi$-plot for
    neutral wetting ($\pca=90^\circ$), $\Par=2$ and $\st\approx0.041$. A
    particle is placed as such that it just touches the undeformed interface. The dashed lines denote the adsorption
    trajectories, the solid lines show the points where the particle touches the
    undeformed interface. The circular
    points depict the stable and the metastable point. The square
    points are related to the snapshots describing the adsorption process in
    the inset. For initial particle orientations of $\pa(t=0)\neq0^\circ$ the
    particle ends in its stable configuration orientated parallel to the
    interface.}
\label{adsorbmpua15h}
\end{figure}

\begin{figure}
\centering
\includegraphics[width=8cm]{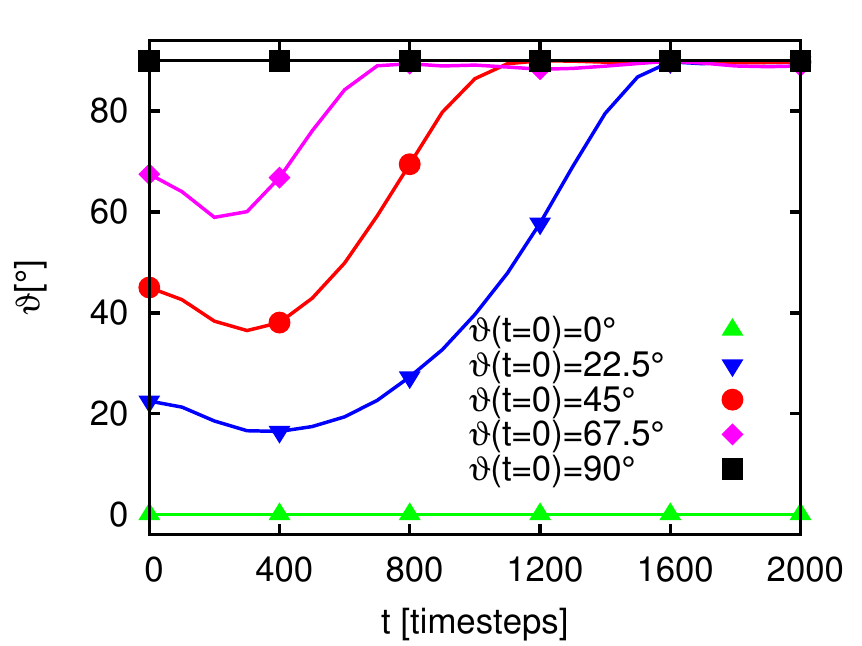}
\caption{Time development of the particle orientation $\pa(t)$ for different
  initial orientations. For $\pa(t=0)\neq0^\circ$ and $\pa(t=0)\neq90^\circ$
  the particle rotates in the `wrong' direction in the first timesteps. The
  time needed to be in the final orientation depends on the initial orientation.}
\label{adsorbplotphit5_1}
\end{figure}

\begin{figure}
\centering
\includegraphics[width=8cm]{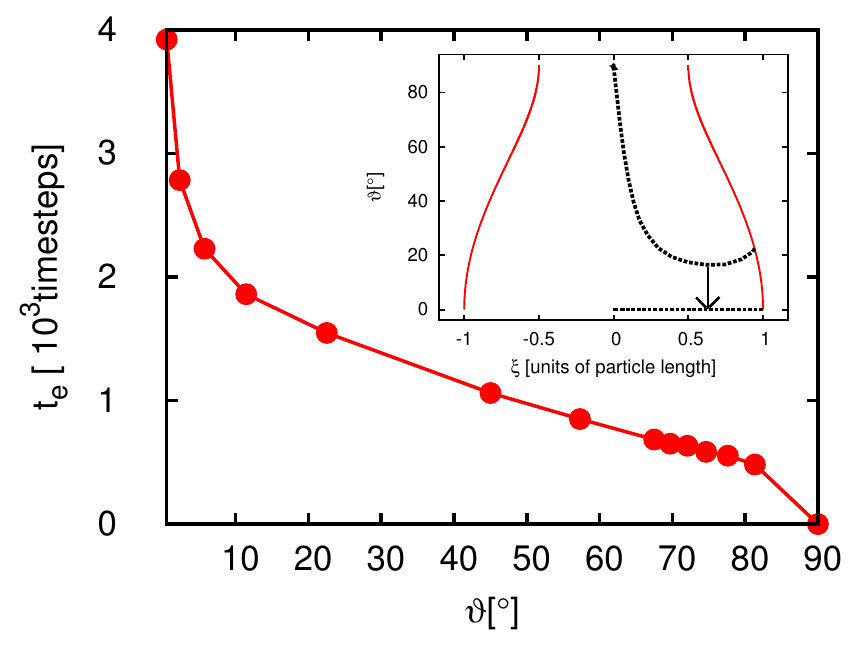}
\caption{Outer plot: Time $t_e$ which the particle needs to reach the final
orientation ($\pa=90^\circ$) for different initial orientation angles
$\pa_0=\pa(t=0)$ from $\pa=0^\circ$ to $\pa=90^\circ$. $t_e$ diverges if
$\pa_0$ approaches $0^\circ$. The reason for the divergence is the approach of
$\pa_0$ to the orientation of the metastable point, as it is shown in the
inset: If the starting angle (middle dashed line) approaches $\pa(t=0)=0$
(lower dashed line) the time required to reach the equilibrium point
diverges.}
\label{ggerreicht0_98}
\end{figure}

In the previous section we demonstrated that there is an additional time
development of the average domain size $L$ for emulsions stabilized by
anisotropic particles. In the following sections we relate this behavior to the
orientational degree of freedom of the particles at the interface. To obtain a
more basic understanding of the additional timescales some simple model systems
are discussed. The simplest possible example is the adsorption of a single
particle at a flat fluid-fluid interface. To characterize the particle
orientation we introduce the angles $\pa$ and
$\Aa$. $\pa$ is the angle between the particle main axis and the $\Doi$-axis,
where the $\Doi$-axis is oriented perpendicular to the flat fluid-fluid
interface. $\Aa$ is the angle between the particle main axis and the
$\dxpi$-axis, where the $\dxpi$-axis is orientated parallel to the interface. $\dip$ is the distance between the particle center and the
undeformed interface in units of the long particle axis.
In this section we consider the case of neutral wetting ($\pca=90^\circ$) and
restrict ourselves to an aspect ratio of $m=\prp/\pro=2$.
The fluid-fluid interaction parameter is set to $g_{br}=0.1$ corresponding to
an interfacial tension of $\st\approx0.041$.
We use a cubic system with 64 lattice nodes in each direction. A wall is placed
at the top and bottom in $\Doi$-direction. Periodic boundary conditions are applied in the
$\dxpi$- and $\dypi$-direction. In order to obtain a flat interface the system
is filled with two equally sized cuboid shaped
lamellae with an interface orthogonal to the $\Doi$-axis. The
lamellae are mainly filled with red and blue fluid, respectively. The initial
majority and minority species are set to $\dens^{\rm maj}=0.7$ and $\dens^{\rm
  min}=0.04$. \\
For this study a
particle is placed so that it just touches the (undeformed) fluid-fluid
interface. This is done for different initial orientations of the particle.
%Fig.~\ref{a15h01}
The inset of Fig.~\ref{adsorbmpua15h} shows snapshots of a typical adsorption process.
%\begin{figure}
%\centering
%\includegraphics[]{}
%\caption{}
%\label{} ma2tbulkadsorb01.4557.out
%\end{figure}
In the beginning the particle is oriented almost orthogonally to the
interface. In the first ca. 2000 timesteps the particle moves towards the
interface without changing its orientation considerably. Then, the particle
rotates and reaches its final orientation after 3600 timesteps.
%Fig.~\ref{adsorbmpp}
The outer plot of Fig.~\ref{adsorbmpua15h} shows a $\pa-\dip$ diagram of the
adsorption. The points where the
particle just touches a flat interface for
the different orientations are marked with solid lines. The dotted lines
indicate the adsorption trajectories. Each black square
is related to one of the snapshots in the inset of Fig.~\ref{adsorbmpua15h}.
Almost all dashed lines end in the upper circle which corresponds to the equilibrium point where the free energy
function has a global minimum. Just the cases with an initial value of
$\pa(t=0)=0^\circ$ end at the metastable point at $\pa=0^\circ$ as shown by the
circle at the bottom of Fig.~\ref{adsorbmpua15h}. This metastable point might not be found in experiments: on the one hand fluctuations will cause a rotation of the particle towards the stable
points and on the other hand, it is impossible to place the particle exactly at $\pa=0^\circ$.\\
Fig.~\ref{adsorbplotphit5_1} and \ref{ggerreicht0_98} depict
the dynamics of the particle adsorption and the influence of the initial
particle orientation $\pa$ with respect to the flat interface.
Fig.~\ref{adsorbplotphit5_1} shows the time development of $\pa$
for  different values of $\pa(t=0)$. For $\pa(t=0)=0^\circ$ and
$\pa(t=0)=90^\circ$ (upper and lower lines) the orientation remains unchanged and the adsorption at
the interface causes only a translational particle movement. The lines for the three other simulation runs %()
start at $\pa(t=0)=22.5^\circ$, $\pa(t=0)=45^\circ$ and
$\pa(t=0)=67.5^\circ$. All of them go in the `wrong' direction during the
first few $10^2$ timesteps and end at $\pa=90^\circ$ corresponding to the stable point, but the time needed for reaching this value
differs. Furthermore, in all cases during the first timesteps, $\pa$ decreases but
then it increases up to this final value. The time $t_e$ the particle needs to
reach the final orientation of $\pa=90^\circ$ depending on $\pa(t=0)$ is shown
in the outer plot of Fig.~\ref{ggerreicht0_98}. 
Due to the discretization of the particle on the lattice,
 its orientation shows small deviations from the theoretical final value. 
Therefore, we measure $t_e$ as the time when the angle reaches $98\%$ of the theoretical final angle. The particle
  oscillates arround this final value but these oscillations are very small and
their magnitude falls below the threshold for the measurement of $t_e$.
%Fig.~\ref{ggerreicht0_98}.
%The time which the particle needs to reach $\pa=90^\circ$ within error limits
%of $98\%$ is defined as $t_e$.
$t_e$ increases with decreasing $\pa$ and diverges for $\pa\rightarrow0$. This
divergence can be understood using the inset of Fig.~\ref{ggerreicht0_98}.
%The approach of the starting angle $\pa(t=0)$ to $\pa=0$ (corresponding to the
%metastable case) which never ends in the stable point 
If the starting angle $\pa(t=0)$ comes closer to $\pa=0^\circ$ (corresponding to the
metastable case where the particle never flips) the capillary forces causing
the particle rotation become smaller and vanish.\\
We have seen that anisotropy of particles causes additional
timescales in the development of
the domain sizes in the emulsions, because of orientational ordering. This timescale is of the
order of $10^{6}$ LB timesteps. In this section we have shown that the
adsorption of a single particle at an interface and its orientational ordering
takes of the order of $10^{3}$ timesteps and depending on the initial particle
orientation towards the interface. We can identify one extra timescale where
the particles rotate towards the interface. This timescale plays a role in the
beginning of the emulsion formation (during droplet formation and droplet growth) when the particles come in contact with the
interfaces. However, this timescale does not yet
explain the full time development.
We require additional model systems to obtain a full understanding of the additional
timescales. Thus, we consider many particles at a flat interface as well as at
a single droplet in the following sections.

\section{Particle ensembles at a flat interface}
After having studied the adsorption of a single particle we discuss the
behavior of a many-particle ensemble at a flat interface. What is the influence
of the hydrodynamic interaction between many particles on the timescales
involved in emulsion formation? For the case of the single-particle adsorption
the particle orientation towards the interface ($\pa$) is an important
parameter. For prolate particles, also the mutual orientation ($\Aa$) of the particles is
important and one has an additional degree of freedom leading to particle
orientational ordering. To characterize the ordering of the particles we use
two order parameters and two correlation functions.

\begin{figure}
\centering
{\includegraphics[width=8 cm]{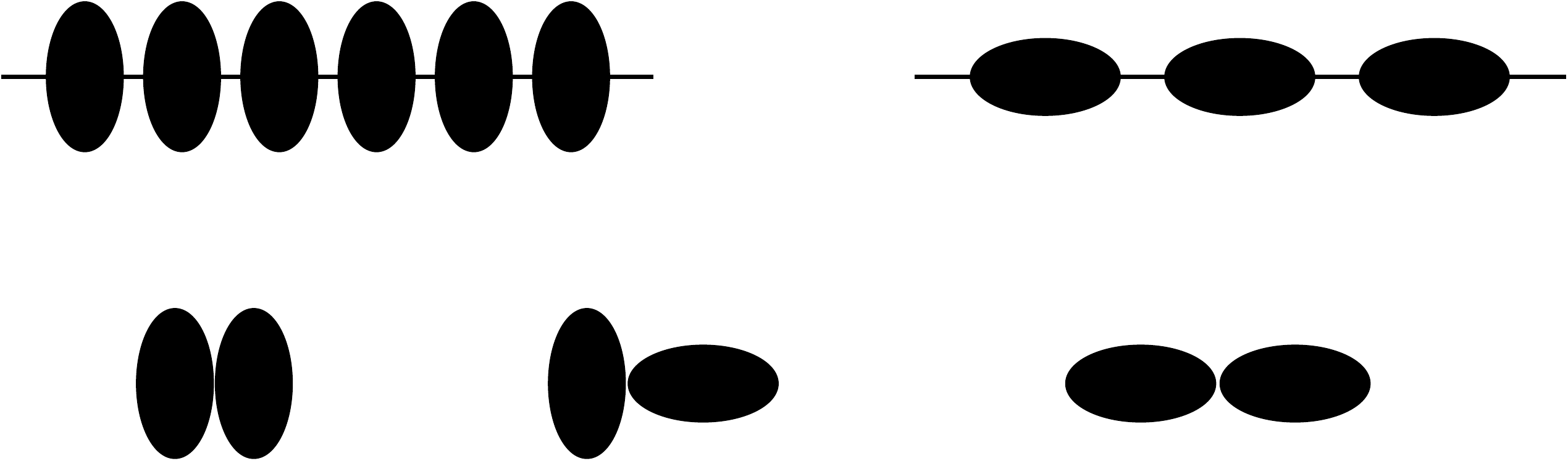}}
\caption{Top right: Sketch of the initial condition for the many-particle system. All
  particles are oriented almost orthogonal to the interface corresponding to the
  initial configuration. Top left: Sketch of the final state. All particles are
  oriented parallel to the interface. Bottom: Different constellations of mutual
  orientation of next neighbors.}
\label{ma2t_initshot2_gesamt}
\end{figure}

Measures for global
ordering effects of the particles are the orientational
order parameters $\opS$ and $\opQ$.
We define the uniaxial order parameter $\opS$~\cite{Kralj1991a,Collings2003a} as
\begin{equation}
  \label{eq:S}
\opS=\frac{1}{2}\left\langle3\cos^2\pa-1\right\rangle
  \mbox{,}
\end{equation}
where $\langle\rangle$ denotes the averaging over particles.
Originally $\opS$ is an order parameter for studying liquid crystals which
indicates the phase transition from the isotropic to the anisotropic/nematic
phase. Here, the parameter $\opS$ is used as a measure for the
orientation of the particle ensemble towards the interface. If all  particles are oriented orthogonal
to the interface we have $\opS=\opS_{\perp{}}=1$ (see top right of
Fig.~\ref{ma2t_initshot2_gesamt}). The orientation of all particles parallel
to the interface leads to $\opS=S_{\parallel{}}=-0.5$ (see top left of Fig.~\ref{ma2t_initshot2_gesamt}).\\The biaxial order parameter $\opQ$~\cite{Collings2003a} is defined as
\begin{equation}
  \label{eq:Q}
\opQ=\frac{3}{2}\left\langle\sin^2\pa\cos(2\Aa)\right\rangle
  \mbox{.}
\end{equation}
%\todo{keep Q in the paper?}
The parameter $\opQ$ is a measure for the mutual orientation of the
particles oriented parallel to the interface. If all particles lying parallel
to the interface are oriented in the same direction it is $\opQ=\opQ_{\rm
  aniso}=1.5$. $\opQ=\opQ_{\rm iso}=0$ means that the particles oriented parallel
to the interface have a two-dimensional isotropic ordering.\\
%The order parameters $\opS$ and $\opQ$ are measures for the global ordering.
The local ordering effects are investigated by using two correlation functions. The
discretized form of the pair correlation function $g(\distpc)$ is defined as
\begin{equation}
  \label{eq:gd}
\Pcf(\distpc)=\frac{1}{2\pi\pcfnorm \pn}\left\langle\sum_{i,j\neq i}\int_{\distpc-\frac{1}{2}}^{\distpc+\frac{1}{2}}\delta(\tilde{\distpc}-\distpc_{ij})d\tilde{\distpc}\right\rangle
  \mbox{,}
\end{equation}
where $\pn$ is the number of particles, $\distpc$ and $\distpc_{ij}$ are the
distance from a reference particle and the distance between the two
%$\distpc$ is the distance between the particle centers in units of $\prp$.
particle centers of particle $i$ and $j$ in units of $\prp$, respectively, and $\pcfnorm$ is a normalization factor
chosen such that $\Pcf(\distpc)\rightarrow1$ for
$r\rightarrow\infty$. $\Pcf(\distpc)$ gives a probability to find a particle
at a distance $\distpc$ from a reference particle. It is a measure for the
ordering of the particle centers and ignores the orientation. As a measure
for the local orientational ordering effects the angular correlation
function~\cite{Cuesta1990a} is defined as (in the discrete form)
\begin{equation}
  \label{eq:g2l}
\acf(\distpc)=\int_{\distpc-\frac{1}{2}}^{\distpc+\frac{1}{2}}\left\langle\cos(2l(\pa(0)-\pa(\distpc))\right\rangle
  \mbox{,}
\end{equation}
with $l=1$ in order to have the appropriate values of $\acf$ for a given
value of $\pa$ discussed below.
$\acf(\distpc)$ gives a measure for the average orientation of particles
at distance $\distpc$ from a reference particle. If the particles at distance
$\distpc$ from the reference particle are all oriented parallel to the
reference particle we have $\acf(\distpc)=1$ (see right and left configuration
\begin{figure}
\centering
{\includegraphics[width=8 cm]{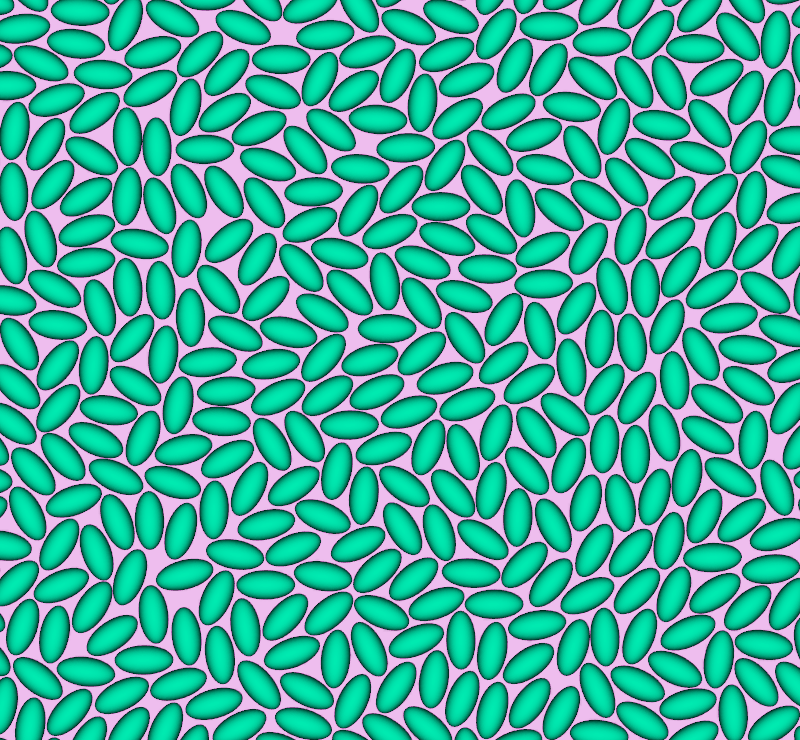}}
\caption{Zoomed snapshot after 10$^4$ timesteps of the state where the
  particles are flipped (Related to top left in Fig.~\ref{ma2t_initshot2_gesamt}).}
\label{ma2t_shotafterfliptop}
\end{figure}
in the bottom of Fig.~\ref{ma2t_initshot2_gesamt}) and an orthogonal orientation leads to
$\acf(\distpc)=-1$ (see central configuration in the bottom of Fig.~\ref{ma2t_initshot2_gesamt}). In the
following we use smoothed versions of $\Pcf$ and $\acf$, where we average over neighboring data points.
%\todo{What are these? Define.}\chng{In order to reduce the noise the noice in
 % the correlation function an average of the neighboring points is used in
 % the following way:
 % $\tilde{\Pcf}(\distpc)=(\Pcf(\distpc-3)+2*\Pcf(\distpc-2)+3*\Pcf(\distpc-1)+4*\Pcf(\distpc)+3*\Pcf(\distpc+1)+2*\Pcf(\distpc+2)+\Pcf(\distpc+3))/16$
 % and
 % $\tilde{\acf}=(\acf(\distpc-3)+2*\acf(\distpc-2)+3*\acf(\distpc-1)+4*\acf(\distpc)+3*\acf(\distpc+1)+2*\acf(\distpc+2)+\acf(\distpc+3))/16$. In
 % the following we skip the tilde in order to keep it simple.}
The flat interface considered in this section is periodic in two dimensions
parallel to the interface and each period has a size of
$\ifa=\ifl^2$, with $\ifl=512=64\prp$. The system is confined by walls
$40$ lattice units distant from the interface in the third dimension. The particle coverage fraction
for $N$ particles adsorbed at the interface is defined as
$\pcf(\dip,\pa)=\frac{N\pifa(\dip,\pa)}{\ifa}$.  $\pifa(\dip,\pa)$ is the area
which the particle would occupy on a hypothetical flat interface and depends on the distance between the particle center
and the undeformed interface and the particle orientation relative to the flat
interface and $\dip$ is the distance between particle center and undeformed interface. In the following we relate the coverage fraction to the case of
$\dip=0$ and $\pa=90^{\circ}$ ($\pcf_I$) or $\pa=90^{\circ}$ ($\pcf_F$) corresponding to the
initial state and the equilibrium state for
$\pca=90^{\circ}$ (see previous section). This leads to $\pcf_I=\frac{NA_{P,I}}{\ifa}$ and
$\pcf_F=\frac{NA_{P,F}}{\ifa}$ with $A_{P,I}=\pi\pro^2$ and $A_{P,F}=\pi\prp\pro$.\\
Initially, the particles are oriented almost orthogonally to the
interface (see top right of Fig.~\ref{ma2t_initshot2_gesamt}). The initial value for the polar angle is chosen as
$\pa\approx0.6^\circ$ for all particles, whereas $\Aa$ and the particle
positions are chosen randomly.
Analogously to the case of the single-particle adsorption the particle flips to an orientation parallel to the interface (see Fig.~\ref{ma2t_shotafterfliptop}). 
%The time development of $\opS$ and $\opQ$ which are measures for the particle
%ordering are shown in Fig.~\ref{OP_2_5_01} for $\pcf=$.
Fig.~\ref{OP_2_5_01} shows the time development of $\opS$ for different values
of $\pcf_I$ ($\pcf_I\approx0.08$ (squares), $\pcf_I\approx0.38$ (circles),
$\pcf_I\approx0.46$ (upward pointing triangles) and $\pcf_I\approx0.52$
(downward pointing triangles)) and the time development of $\opQ$ for $\pcf_I\approx0.38$ (diamonds).
%$\opS$ starts at $\pcf=....$.
\begin{figure}
%\centering
%\includegraphics[width=3.6 cm]{OP01vglpx.pdf}
  \subfigure[]
%  {\label{OP_2_5_01}\includegraphics[width=3.6 cm, height=3.6 cm]{OPSQ02x.pdf}}
  {\label{OP_2_5_01}\includegraphics[width=8 cm]{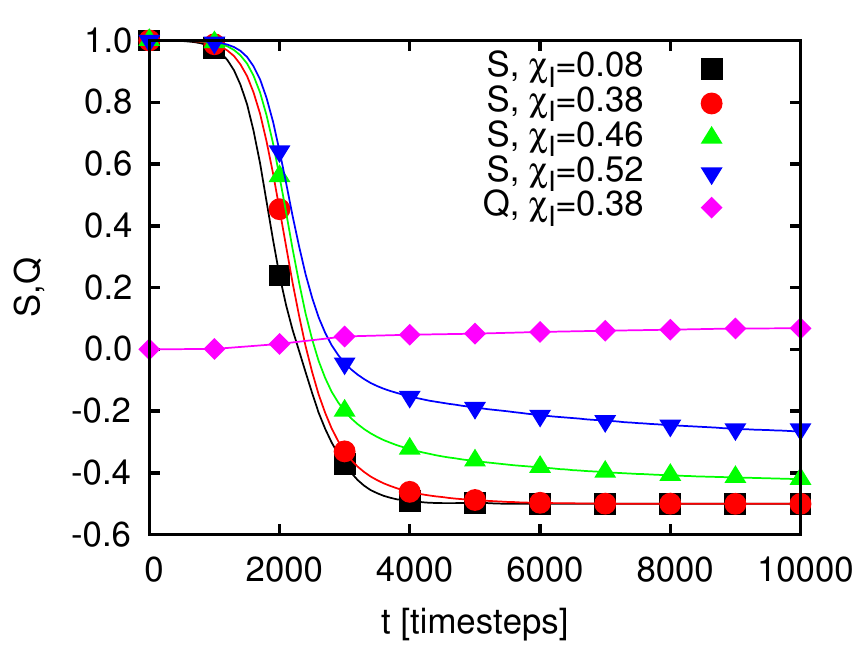}}

\qquad
  \subfigure[]
  {\label{SD_sigma}\includegraphics[width=8 cm]{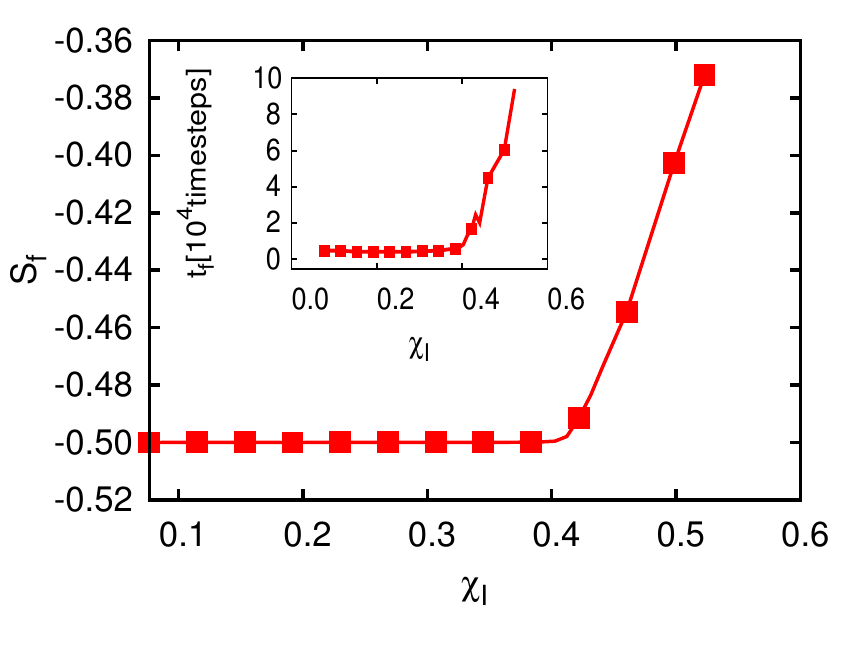}}
%  {\label{SD_sigma}\includegraphics[width=8 cm]{Sf_100000ex.pdf}}
\caption{(a) Time development of the two order parameters $\opS(t)$ and
  $\opQ(t)$ (see \eqnref{S} and \eqnref{Q}) for $m=2$, $\pca=90^\circ$, $\sigma\approx0.041$.
   $\opQ(t)$ is shown for
  a single value of $\pcf_I$ only since it stays at a value of approximately
  0 for all $\pcf_I$. $\opS(t)$ is shown for different values of
  $\pcf_I$. In case of highly packed interfaces, i.e. for large values of
  $\pcf_I$, not all particles are able to fully align with the interface. For
  larger values of $\pcf_I$ $\opS$ needs a longer time to get into the
  equilibrium than shown here.
%\todo{put $\opS(t)$ for different values of $N$ in the plot, also for
%  $\opS_{final}\neq-0.5$. put also more correlation functions in the
%  corresponding plot.}
%800 rot 0.690291354549
%100 green 0.0862864193186
%1000 blue 0.862864193186
%1200 purple 1.03543703182
(b) Outer plot: The final values of the order parameter $\opS$ are plotted
  for different particle densities $\pcf_I$. As shown in
  \ref{OP_2_5_01} a transition from a fully ordered to a disordered state can be found at a critical value of
  $\pcf_{I,C}\approx0.42$. Inset: the time the order
  parameter $S$ (defined in Eq.~\ref{eq:S}) requires to reach the 
  final value (time which particles need to flip). For small values of
  $\pcf_I$ $t_f$ is independent of $\pcf_I$ but above a critical value of $\pcf_I=\pcf_{I,C}$
  $t_f$ increases with increasing $\pcf$ by almost one order of magnitude. 
%\dn -8\{{k}1o
}
%\label{OP_2_5_01}
\end{figure}
The parameter $\opQ$ starts at 0 and
ends at a small value ($\opQ_{\rm final}\approx0.05\ll \opQ_{\rm aniso}$) far away from the
value of total ordering. A similar behavior is found for all values of $\pcf_I$.
Fig.~\ref{ma2t_shotafterfliptop} shows that there are smaller domains where
particles are oriented in the same direction. But every domain has a
different preferred particle direction which might lead to small but still finite values of $\opQ$.
Another reason for this effect is the finite system
size and finite particle number which change the parameter as follows~\cite{Cuesta1990a}:
\begin{equation}
  \label{eq:Qfinite}
\opQ=\opQ_{\infty}+O\left(\frac{1}{\sqrt{N}}\right)
  \mbox{.}
\end{equation}
$\opQ_{\infty}$ is the value of the biaxial order parameter that the
corresponding system with an infinite amount of particles would have.

The parameter $\opS$ starts for all values of $\pcf_I$ with a value of $\opS_{\perp{}}=1$,
corresponding to the initial configuration. For lower values of $\pcf_I$ the
parameter $\opS$ reaches $\opS_{\parallel{}}=-0.5$, corresponding to the
case that the particles flip completely.
For higher values of $\pcf_I$ the final value of the parameter is $\opS_{\rm final}>\opS_{\parallel{}}=-0.5$. This corresponds to the case where some particles cannot flip completely to the equilibrium orientation because there
is insufficient space.
%Fig.~\ref{SD_sigma} shows the final value 
%\todo{compare phase transition in \cite{Cuesta1990a} for $m=2$ with own
%  results.}
The final values of $\opS$ (obtained after $10^5$ timesteps) are shown in the
outer plot of Fig.~\ref{SD_sigma} as a function of $\pcf_I$. We find a
transition point at $\pcf_{I,C}\approx0.42$ corresponding to
$\pcf_{F,C}\approx0.84$. If all particles are oriented parallel to
the interface the system corresponds practically to a two dimensional system of
ellipses. However, the value of $\pcf_{I,C}$ found is below the value of the
closest packing density for a two-dimensional system of ellipses with
$\Par=2$, which is $\pcf_{F,\rm max}\approx0.91$.
Such a system was also studied in Ref.~\cite{Cuesta1990a} with Monte
Carlo simulations. For the case of an ellipse with an aspect ratio $\Par_{2d}=2$ a transition point of
$\pcf_{\rm 2dmc}\approx0.78$ from isotropy to a solid phase was found. The
solid phase describes a state where the particle centers as well as the
orientations are ordered. We do not reach the limit of the solid phase.
This suggests that hydrodynamic interactions and absence confinement in the
third dimension still play a dominant role. The biaxial order parameter in the
MC system grows up to
$Q\approx1$ (see Fig. 11 in Ref.~\cite{Cuesta1990a}) corresponding to a global anisotropic state with a quite high
degree of ordering for $\pcf_F>\pcf_{\rm 2dmc}$. This effect is not observed in
our system. The reason
for this difference is the method used to reach this state. A
two-dimensional system of ellipses was studied in  Ref.~\cite{Cuesta1990a} wheres
we simulated three-dimensional ellipsoids which form an effective
two-dimensional system by flipping to the interface.\\
We can see that in the many-particle system and for small and moderate $\pcf_I$ about $10^3$ timesteps are required for the particles
to flip which is the same order of magnitude as in the case of the single
particle adsorption for small values of $\pcf_I$.
%The fact of having many particles does not influence the timescale of particle flipping. 
The inset in Fig.~\ref{SD_sigma} shows the time the order parameter $\opS$ needs to
reach its final value. This corresponds to the time required for the whole particle
ensemble to be flipped completely ($\pcf_I<\pcf_c$) or to reach the
semi-flipped state for $\pcf_I>\pcf_c$. For $\pcf_I<0.38$ $t_f$ stays almost
constant at about 4500 timesteps. In this regime the distance between the
particles is sufficient so that the influence of hydrodynamic interactions
on the flipping behavior can be neglected.
For higher values it increases very sharply and
hydrodynamic interactions between the particles must not be neglected
anymore. Furthermore, the time needed to flip completely for the very dense
systems (jammed state) is about one order of magnitude larger.
% \todo{add it later in the discussion of ts in emulsions}.

The biaxial order parameter does not show any global ordering but the
snapshot in Fig.~\ref{ma2t_shotafterfliptop} shows
some local ordering effects. Hence, we need other ways to characterize the local
ordering effects and utilize the two local correlation
functions $\Pcf(\distpc)$ and $\acf(\distpc)$ defined above.
The particles have a contact angle of $90^\circ$, so there are no capillary
interactions between them in the final state when all of them have flipped
completely and the system has reached an equilibrium. However, there are dipolar
interface deformations and thus the interactions during the flipping process of the particles and for
$\pcf>\pcf_c$ which causes capillary interactions at this time. After flipping
there are still some capillary waves going through the system, leading to
interactions between the particles.
The pair correlation function
\begin{figure}
  \centering
  \subfigure[]
%  {\label{Grm5_01}\includegraphics[width=3.6 cm, height=3.6 cm]{g01.pdf}}
  {\label{Grm5_01}\includegraphics[width=8 cm]{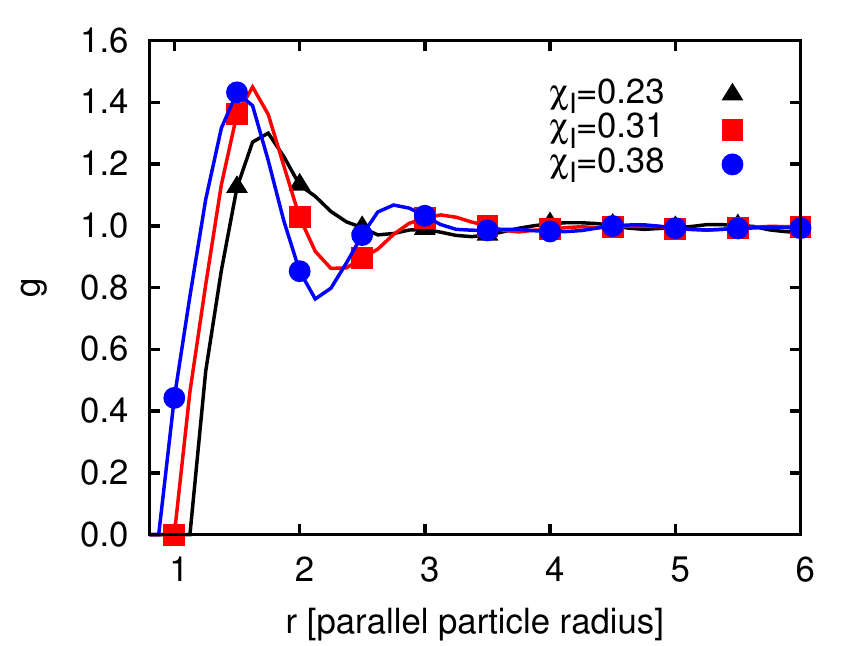}}
\qquad
  \subfigure[]
  {\label{G2thm2_01}\includegraphics[width=8 cm]{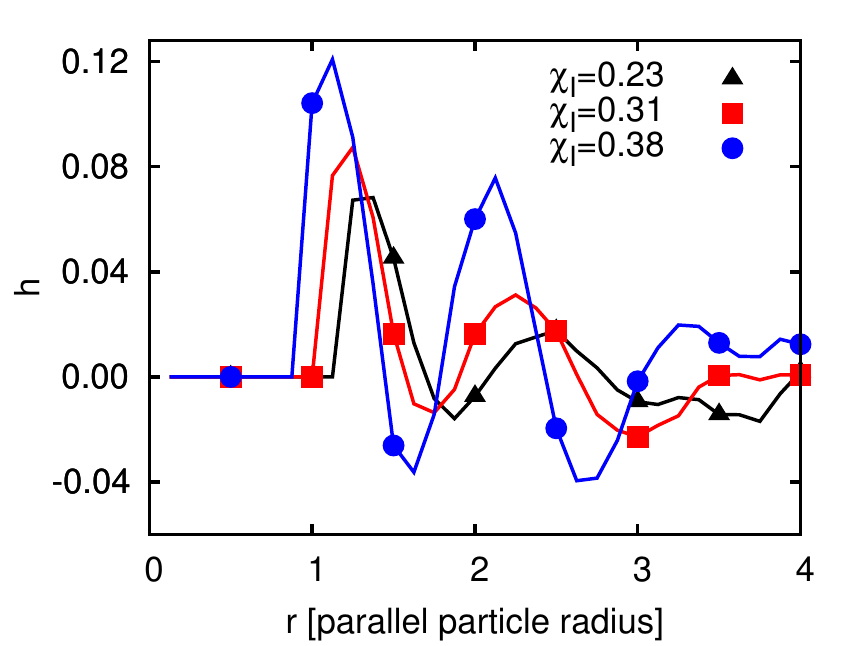}}
  \caption{(a) Pair correlation function $\Pcf(\distpc)$ (defined in \eqnref{gd}) (b)
    orientation correlation function $h(\distpc)$ (defined in
    \eqnref{g2l}). In both cases the ordering increases with increasing $\pcf_I$.}
\end{figure}
$\Pcf(\distpc)$ is shown in Fig.~\ref{Grm5_01} for three different values of
$\pcf_I$ ($\pcf_I\approx0.23$, $\pcf_I\approx0.31$ and
$\pcf_I\approx0.38$) after $10^5$ timesteps.
The first peak is pronounced in all three cases. The distance $\distpc$
of this peak decreases for increasing $\pcf_I$ as well as the degree of
ordering. For the highest $\pcf_I$ a depletion region leading to a minimum
after the peak is pronounced.
To obtain a measure of the local orientational ordering effects
we investigate the orientational correlation
function $h(\distpc)$ as shown in Fig.~\ref{G2thm2_01} for the same 3 values
of $\pcf_I$.
%\todo{add plot for$\pcf_I\approx0.086$ and discuss differences in peaks etc.}.
%\todo{keep 23 in? If yes: what do we learn and more lines?}
The first two positive peaks and the first negative peak can be explained with the
drawings in the bottom of Fig.~\ref{ma2t_initshot2_gesamt}. The first positive peak is due to a
side-to-side alignment of two particles. Fig.~\ref{ma2t_shotafterfliptop}
shows several domains of side-to-side alignment. The first negative peak comes from an
alignment where the particles are oriented perpendicular to each other and
the second positive peak comes from a tip-to-tip alignment or second nearest
neighbors of side-to-side orientation. The degree of translational and
orientational ordering increases with increasing $\pcf_I$.\\
\begin{figure}
\centering
\includegraphics[width=8cm]{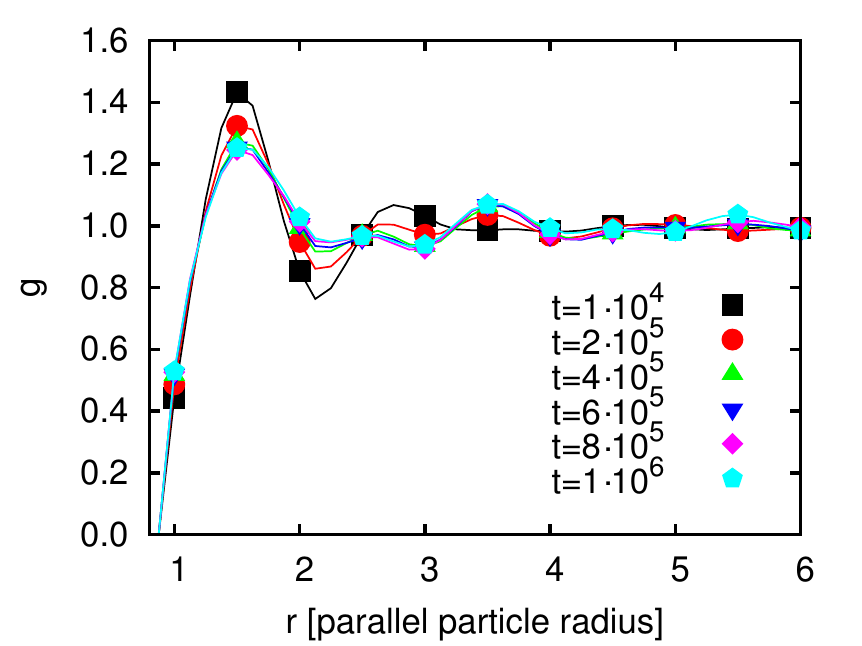}
\caption{Time development of $\Pcf(\distpc)$ for $\pcf_I\approx0.38$. The
  second peak is more pronounced at later timesteps. The particles reorder and
  the ordering increases. The reordering process is almost done after $4\cdot10^5$ timesteps.}
\label{Grm5_01tdev}
\end{figure}
After having discussed the correlation functions we investigate the time
development of $g(r)$ in order to understand
the time development of the average domain size $\ds$. 
%The adsorption and rotation of single particles towards the interface is not sufficient to explain the long time behavior of $\ds$. Therefore, we now investigate the collective alignment of particles at a fluid interface.
Fig.~\ref{Grm5_01tdev} shows $\Pcf(\distpc)$ at different times between
10$^4$ and 10$^6$ timesteps. The first peak decreases but at later times the
following peaks are more pronounced. Thus, the degree of ordering
increases. After $4\cdot10^5$ timesteps this development has almost come to an end.
The reason for this remaining development
is the particle reordering. The particles form domains where they align
parallel to each other. These domains become larger with time.

In this section we have shown
shows that the presence of many particles at an interface leads
to two additional timescales in the reordering. The first one is the rotation
of the particle towards the interface. The particle rotates towards its final
orientation parallel to the interface. For lower values of $\pcf_I$ this
process does not depend on $\pcf_I$ and is not different from the single
particle adsorption. For larger values of $\pcf_I$ the time needed to come to
its final orientation increases. Hydrodynamic as well as excluded volume
effects become more important. Above a critical value not every particle reaches
its `final' orientation. 
The reordering of $h$ (corresponding to $g$ in Fig.~\ref{Grm5_01tdev}) can also
be observed. The first 2 peaks get more pronounced after several $10^5$
timesteps as compared to the state after 10$^4$ timesteps shown in Fig.~\ref{G2thm2_01}.
%This instant of time is chosen as a starting point in order
%to avoid particle flipping effects from the first time.

%\todo{Text richtig einbauen:}
%We can identify three timescales of this development. The first
%timescale is the forming of the domains and the particle adsorption at the
%interface. For the case of spheres, everything is done after this
%timescale. For the case of anisotropic particles, two more timescales can be found: the second timescale is caused by the rotation towards the
%interface. The mutual ordering of the particles orientated parallel to the
%interface causes the third timescale. If the interface is curved, a change of
%the mutual orientation of particles (both oriented parallel to the interface)
%changes the interface shape. The interfaces of Pickering emulsions and bijels
%are generally curved.
% $\,$

%\input{vtdropletadsorption}
\section{Particle ensembles at a spherical interface}
In the previous chapter the behavior of particle ensembles at a flat interface
was discussed. However, in emulsions the interfaces are generally not flat.
Pickering emulsions usually have (approximately) spherical droplets and a bijel
has an even more complicated structure of curved interface. The simplest
realization of a curved interface is a single droplet and as such is studied in
this section.

The simulated system is periodic and each period has a size of $L_S=256$
lattice units. The droplet radius and the number of adsorbed particles are
chosen to be $R_D=0.6L_S\approx76.8$ and 600, respectively. In the beginning of
the simulation the particles are placed orthogonal to the local interface
tangential plane. As we have seen already for the case of flat interfaces the
particles flip to an orientation parallel to this tangential plane. This state
is shown in Fig.~\ref{snapshot_droplet} after $2\cdot10^5$ timesteps.
\begin{figure}
\centering
\includegraphics[width=8cm]{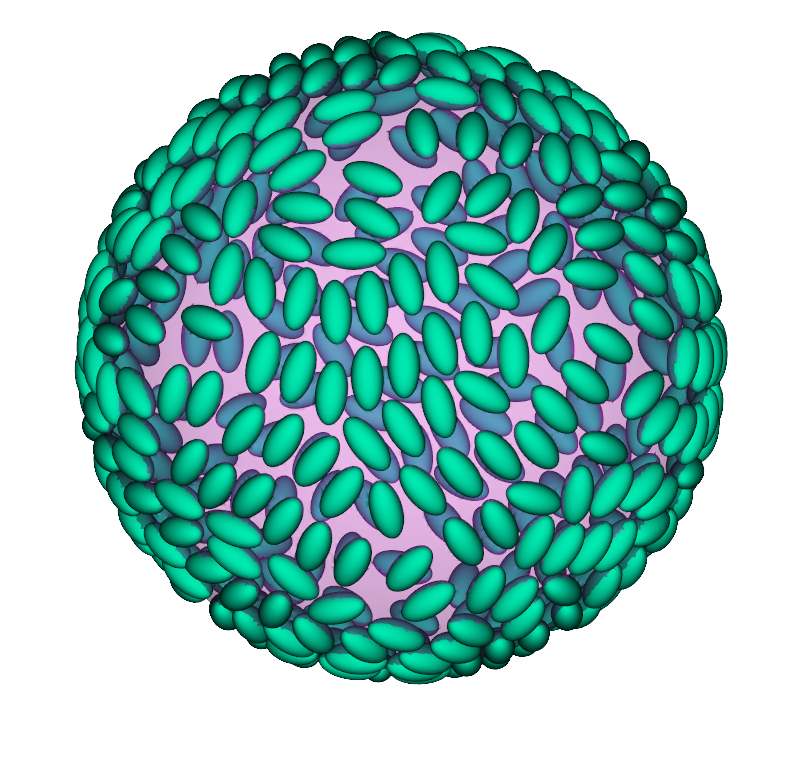}
\caption{Snapshot of a particle ensemble at a spherical interface after $2\cdot10^5$ timesteps.}
\label{snapshot_droplet}
\end{figure}
A preliminary comparison between flat and spherical interfaces has already been
given in our previous contribution~\cite{Krueger2012c}. The time development of
$\opS$ is shown in Fig.~11(a) in Ref.~\cite{Krueger2012c}. It has been found
that the influence of the interface curvature on the flipping process is larger
than the influence of the particle coverage. The time needed for the particles
to flip is about a factor two smaller in the case of the curved interface.

\begin{figure}
\centering
\includegraphics[width=8cm]{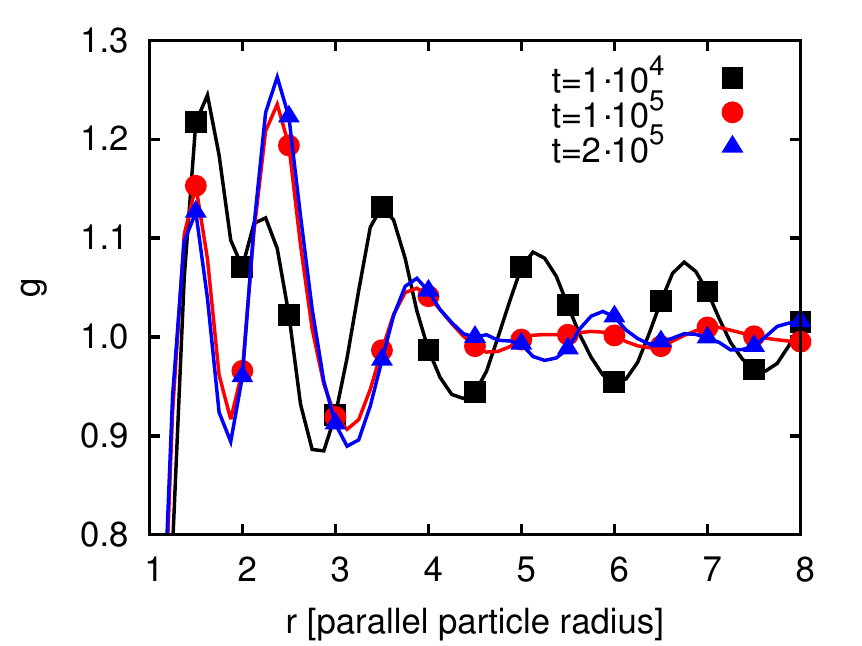}
\caption{Time development of the order parameter $g(r)$ for particles adsorbed
at a spherical interface.}
\label{Grm5_droplet}
\end{figure}
Here, we investigate the particle correlation function (see \eqnref{gd}) for
the particle ensemble. Fig.~\ref{Grm5_droplet} shows $\Pcf$ for
$\pcf_I\approx0.27$ at three different times. After 10$^4$ timesteps it is
still close to the correlation function of the initial condition. After 10$^5$
timesteps some changes can be seen. The first peak is reduced but the second
peak is more pronounced. There is no substantial change between $1\cdot 10^5$
and $2\cdot 10^5$ timesteps. Compared to the state at 10$^5$ timesteps the
correlation function shows pronounced peaks at longer distances from the
particle (about $6 R_p$). The particles mostly reorder during the first 10$^5$
timesteps since at later times only minor changes in the particle order can be
observed.  Similar to the case of flat interfaces that was discussed in the
previous section, the particle ensemble forms domains where the particles are
ordered in a nematic fashion.  The peaks in the correlation function are more
pronounced in the case of droplets than in the case of a flat interface.  The
reason is given by the capillary interactions between the particles which are
much stronger in the case of curved interfaces. In particular, non-zero
capillary interactions persist between spheroids even in the case of neutrally
wetting particles.

The time development of $\Pcf$ at the droplet as discussed in this section
differs from the behavior in the case of a flat interface.  For the droplet,
$\Pcf$ arrives at its final structure after about 10$^5$ timesteps whereas at
the flat interface about four times more as many steps are required.  In
addition, for flat interfaces, $\Pcf$ only shows one or two peaks (depending on
$\pcf_I$), while for the particle covered droplet five peaks are found due to a
larger range of ordering of the particles. This is a result of the stronger
capillary interactions between the particles due to the interface curvature.

We can understand one of the additional timescales with the behavior of the
ellipsoidal particles at a single droplet. The particles reorder and it can be
shown that this leads to a small deviation of the shape of the droplet which is
(almost) exactly spherical in the beginning~\cite{Kim2008a}. A change of the
interface shape caused by reordering of anisotropic particles leads to a change
of $L(t)$.  The reordering of particle ensembles at flat as well as spherical
interfaces takes of the order of $10^5$ timesteps. This reordering takes place
in idealized systems with constant interfaces which do not change their shape
considerably. In real emulsions, however, the interface geometry changes
substantially during their formation. For example, two droplets of a Pickering
emulsion can coalesce. After this unification the particle ordering starts
a new. This explains the fact that the additional timescale we find in our
emulsions is of the order of several $10^6$ timesteps.

\section{Conclusion}
In this article we have investigated the dynamics of the formation of Pickering
emulsions and bijels stabilized by ellipsoidal particles. In contrast to
emulsions stabilized by spherical particles, spheroids cause the average time
dependent droplet or domain size to slowly {\it decrease} even after very long
simulation times corresponding to several million simulation timesteps. The
additional timescales related to this effect have been investigated by detailed
studies of simple model systems. At first, the adsorption of single ellipsoidal
particles was shown to happen on a comparably short timescale ($\approx 10^4$
timesteps). Second, many particle ensembles at flat interfaces, however, might
require substantially more time in case of sufficiently densely packed
interfaces. Here, local reordering effects induced by hydrodynamic
interactions and interface rearrangements prevent the system from attaining a steady
state and add a further timescale to the emulsion formation ($\approx 10^5$
timesteps). Third, this reordering is pronounced in the case of curved
interfaces, where the movement of the particles leads to interface deformations
and capillary interactions. During the formation of an emulsion, droplets might
coalesce (Pickering emulsions) or domains might merge (bijels). After such an
event the particles at the interface have to rearrange in order to adhere to
the new interface structure. Due to this, the local reordering is practically
being ``restarted'' leading to an overall increase of the interfacial area on a
timescale of at least several $10^6$ timesteps. With the nanoscale
resolution chosen above, this corresponds to physical
times of the order of $10^{-5}s$.

Our findings provide relevant insight in the dynamics of emulsion formation
which is generally difficult to investigate experimentally due to the required
high temporal resolution of the measurement method and limited optical
transparency of the experimental system. It is well known that in general
particle-stabilized emulsions are not thermodynamically stable and therefore the
involved fluids will always phase separate -- even if this might take several months.
Anisotropic particles, however, provide properties which might allow the
generation of emulsions that are stable on substantially longer timescales.
This is due to the continuous reordering of the particles at liquid
interfaces which leads to an increase in interfacial area and as such
counteracts the thermodynamically driven reduction of interface area.

\begin{acknowledgments}
Financial support is greatly acknowledged from
NWO/STW (Vidi grant 10787 of J.~Harting) and FOM/Shell IPP (09iPOG14 -
``Detection and guidance of nanoparticles for enhanced oil recovery'').
We thank the J\"ulich Supercomputing Centre, SARA Amsterdam, and HLRS
Stuttgart for computing resources. J. de Graaf, M. Dijkstra, and R. van Roij are kindly acknowledged for
fruitful discussions.
\end{acknowledgments}
%\bibliographystyle{unsrtnat}
%\bibliography{bibliography}

\end{document}